\documentclass[aps,twocolumn,showpacs,amsmath,amsfonts,amssymb]{revtex4}
\usepackage{graphicx}
\usepackage{epsf}

\begin{document}
\title{Diffraction limit of the sub-Planck structures}

\author{Raman
Sharma$^{\mathrm{1}}$\footnote{e-mail:ramansharma@cse.iitb.ac.in},
Suranjana Ghosh$^{\mathrm{2}}$\footnote{e-mail:sghosh@iitp.ac.in},
Utpal Roy$^{\mathrm{2}}$\footnote{e-mail:uroy@iitp.ac.in}, and
Prasanta K.
Panigrahi$^{\mathrm{3}}$\footnote{e-mail:pprasanta@iiserkol.ac.in}}

\affiliation{$^{\mathrm{1}}$Indian Institute of Technology Bombay,
Powai, Mumbai - 400076, India
\\ $^{\mathrm{2}}$ Indian Institute of Technology Patna, Patliputra Colony, Patna 800013, India
\\$^{\mathrm{3}}$ Indian Institute
of Science Education and Research Kolkata, BCKV Campus Main
Office, Mohanpur - 741252, India}

\begin{abstract}
The orthogonality of cat and displaced cat states, underlying
Heisenberg limited measurement in quantum metrology, is studied in
the limit of large number of states. The asymptotic expression for
the corresponding state overlap function, controlled by the
sub-Planck structures arising from phase space interference, is
obtained exactly. The validity of large phase space support, in
which context the asymptotic limit is achieved, is discussed in
detail. For large number of coherent states, uniformly located on
a circle, it identically matches with the diffraction pattern for
a circular ring with uniform angular source strength. This is in
accordance with the van Cittert-Zernike theorem, where the overlap
function, similar to the mutual coherence function matches with a
diffraction pattern.
\end{abstract}

\maketitle

\section{Introduction}

Cat states and their generalizations are known to achieve
Heisenberg limited sensitivity in estimation of parameters like
coordinate/momentum displacements and phase space rotations
\cite{Zurek}. A criterion to distinguish quantum states without
classical counterparts, from those without the same, are studied
in \cite{Vogel,Richter}. For these non-classical states, subtle
interference effects in the phase space \cite{Schleich Wheeler}
lead to sub-Planck structures in their Wigner functions, which in
turn allow precision measurement of quantum parameters, bettering
the standard quantum limit. Sub-Planck structures in different
physical systems have been recently investigated
\cite{GSAg,Jay,Toscano,ghosh,Roy,sghosh1,sghosh2}. It has been
demonstrated \cite{Toscano,ghosh,Roy} that the sensitivity of the
state used in quantum metrology is directly related to the area of
the sub-Planck structures: $ \rho =\frac{\hbar^2}{A}$, with $A$
being the action of the effective support of the Wigner function.
The interference in phase space is a pure quantum phenomenon,
arising due to the fact that these states are superposition of the
coherent states (CSs), which themselves are classical. The
increase in the number of interfering coherent states in the phase
space is akin to emergence of diffraction in classical optics,
when the number of interfering sources becomes large.

Here, we analyze this diffraction limit of the sub-Planck
structures and find an exact asymptotic value of the displacement
sensitivity. With the assumption of large phase-space support for
the estimating state and smallness of the quantum parameters to be
estimated, it is found that the asymptotic limit of the
sensitivity reaches $|\delta| = \frac{C}{2 |\alpha|}$, where $C$
is the first root of $J_0$, the $0^{th}$ order Bessel function. We
explicitly show that this assumption is adequate to adapt
realistic values of the physical parameters; \emph{i.e.}, the
average photon number and the number of superposed CSs. The
numerical analysis depicts how the asymptotic limit of exact
overlap function (OF) reaches to the $0^{th}$ order Bessel
function for higher order mesoscopic superpositions. This limiting
behaviour in the phase space interference is found to be the exact
analog of the van Cittert-Zernike theorem \cite{Wolf}, relating
the mutual coherence in classical optics to diffraction.

\section{Results and Discussions}

Cat states and their generalizations play a significant role in
quantum optics and quantum computation \cite{Milburn}. A number of
experimental schemes exist to produce cat states in laboratory
conditions \cite{Haroche}. These ``pointer states''\cite{pointer}
often naturally manifest, when suitable quantum systems are
coupled with decohering environment. It has been observed that the
robustness of these states, made out of classical CSs, is a result
of ``quantum Darwinism'' \cite{Darwin}. We consider a single
oscillator, with the CS being an eigen state of $a$:
$a|\alpha\rangle = \alpha |\alpha\rangle$, with annihilation and
creation operator $a$ and $a^{\dagger}$ : $[a,a^{\dagger}]=1$.

The generalized cat state is composed of CSs, equally phase
displaced on a circle:
\begin{equation}\label{cat}
 |cat_{n,\alpha} \rangle = \frac{1}{\sqrt{n}} \sum_{j=1}^{n} |e^{\frac{\iota 2 \pi j}{n} } \alpha \rangle = \frac{1}{\sqrt{n}} \sum_{j=1}^{n} D(e^{\frac{\iota 2 \pi j}{n} } \alpha )|0\rangle \text{,}
\end{equation}
where, $|\alpha\rangle = D(\alpha) |0\rangle$, with the displacement operator, $D(\alpha)=e^{\alpha a^{\dagger} -
\alpha^{\star} a}$ and $a|0\rangle=0$.

\begin{figure*}[htbp]
\centering
\includegraphics[width=4in]{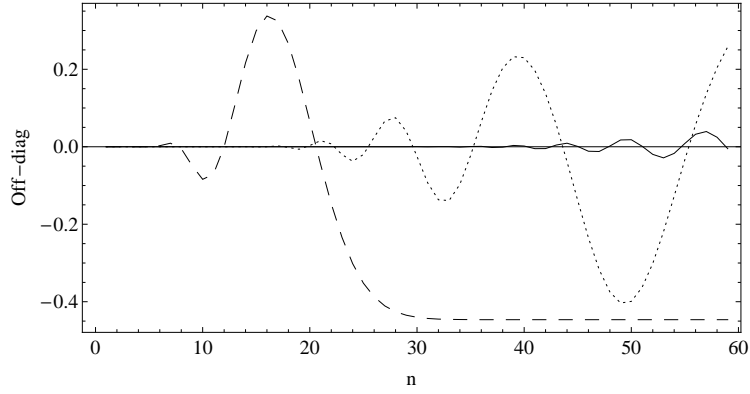}
\caption{Contribution of the off-diagonal terms ($j\neq k$) in
Eq.~(\ref{over1}) or (\ref{over2})) for different phase space
supports: $\alpha=4$ (dashed line), $\alpha=10$ (dotted line),
$\alpha=20$ (solid line), for an arbitrary fixed value of
$\delta=0.2$.}\label{offdiag}
\end{figure*}

The displacements in the coordinate and momenta can be realized
through an appropriately displaced cat state \cite{Toscano}:
\(|cat_{n,\alpha}^{\delta}\rangle=D(\delta)|cat_{n,\alpha}\rangle\).
For checking the sensitivity of the estimating state
$|cat_{n,\alpha} \rangle$, one computes the overlap of the same
with the displaced state and studies the orthogonality conditions,
\begin{widetext}
\begin{eqnarray}
\langle cat_{n,\alpha}|cat_{n,\alpha}^{\delta}\rangle
&=& \frac{1}{n} \sum_{j=1}^n \sum_{k=1}^n \langle 0|D(e^{\frac{i 2 \pi j}{n} }
\alpha )^{\dagger} D(\delta) D(e^{\frac{i 2 \pi k}{n} } \alpha ) |0\rangle \nonumber \\
&=& \frac{1}{n} \sum_{j=1}^n \sum_{k=1}^n (e^{i Im(\delta
\alpha^{\star} ( e^{- \frac{i 2 \pi j}{n}}+e^{- \frac{i 2 \pi
k}{n}}) + |\alpha|^2 e^{- \frac{i 2 \pi (k-j)}{n}})}) (e^{-
\frac{1}{2} |\delta + \alpha (e^{- \frac{i 2 \pi k}{n}} - e^{-
\frac{i 2 \pi j}{n}})|^2})
\text{.}\label{over1}\\
&=&\frac{1}{n} \sum_{j=1}^n \sum_{k=1}^n e^{i(2r
cos(\frac{\pi(j-k)}{n})\;sin(\theta-\frac{\pi(j+k)}{n})+|\alpha|^2\;sin(\frac{2\pi(j-k)}{n}))}\nonumber\\
&&\;\;\;\;\;\;\;\;\;\;\times
e^{-\frac{1}{2}(|\delta|^2+2|\alpha|^2(1-cos(\frac{2\pi(j-k)}{n}))+4r\;sin(\frac{\pi(j-k)}{n})sin(\theta-\frac{\pi(j+k)}{n}))}\label{over2}
\end{eqnarray}
\end{widetext}

where, $r = |\alpha||\delta|$ and $\theta = (\theta_\delta
-\theta_\alpha)$ with $\alpha = |\alpha|e^{i \theta_\alpha}$ and
$\delta = |\delta|e^{i \theta_\delta}$.

\begin{figure*}[htbp]
\centering
\includegraphics[scale=0.5]{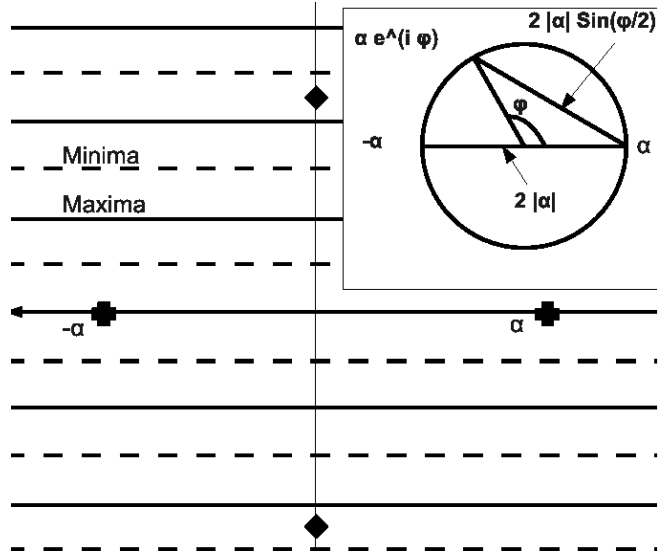}
\caption{\textbf{Analogy between two-source interference:} The
solid lines show the maximum and dashed lines show the minimum
intensity values. The crosses are the positions of coherent state
and diamonds show the equivalent positions of sources of light
which will produce the same pattern at a distance. The inset shows
the equivalent position of sources for the state
$|cat2_{\alpha,\phi}\rangle = \frac{|\alpha\rangle+|\alpha
e^{\iota \phi}\rangle}{2}$} \label{figure}
\end{figure*}

The entire contribution of the OF mainly originates from the
adjacent components of the original and displaced cat states, {\it
i.e.}, $j\sim k$. Therefore, $|j-k|<<n$,
$cos(\pi(j-k)/n)\rightarrow 1$ and $sin(\pi(j-k)/n)\rightarrow 0$.
Then Eq.~(\ref{over2}) takes the simpler form
\begin{equation}
\langle cat_{n,\alpha}|cat_{n,\alpha}^{\delta}\rangle =
\frac{e^{-\frac{1}{2}|\delta|^2}}{n} \sum_{j=1}^n \sum_{k=1}^n
cos\left[2r sin(\theta-\frac{\pi(j+k)}{n})\right].\label{over4}
\end{equation}
Apparently, the off-diagonal terms in the above expression are
only significant for higher order superpositions ($n>>1$). Now,
with the assumption of large phase-space support for the
estimating state and smallness of quantum parameters to be
estimated, one can neglect the off-diagonal terms and obtain
\begin{eqnarray}
\langle cat_{n,\alpha}|cat_{n,\alpha}^{\delta}\rangle &\approx&
\frac{e^{-\frac{1}{2}|\delta|^2}}{n} \sum_{j=1}^{n} cos\left[2r
  sin(\theta-\frac{2\pi j}{n})\right] \nonumber\\
  &\approx& \frac{1}{n} \sum_{j=1}^{n} cos\left[2r sin(\theta-\frac{2\pi
j}{n})\right]\text{.}
\end{eqnarray}
The phase space of the generalized cat state of Eq.~\ref{cat} is
composed of `$n$' CSs, equally placed in a circle of radius
$|\alpha|$, where large phase space support means the large
magnitude of $|\alpha|$.

In this context, we must emphasize more on the domain of validity
of the above assumption. Fig.~\ref{offdiag} depicts the
contribution of the off-diagonal terms of Eq.~(\ref{over1}) or
Eq.~(\ref{over2}). It becomes significant only after very large
value of '$n$' or when $n/\alpha$ ratio is approximately two for a
fixed phase space support. Increasing $n$ or producing a higher
order superposition is quite difficult in experiments, as it
requires a large nonlinearity of the medium. On the contrary, the
absolute value of $\alpha$ is directly related to the average
photon number of the coherent state, which can be manipulated by
controlling the laser beam. Hence, the allowed maximum order of
mesoscopic superposition (`$n$') for a given $\alpha$, conforming
our assumption, is sufficiently large in reality.

It needs to be mentioned that the state overlap depends only on
$\delta \alpha^{\star}$, which leads to the conclusion that the
sensitivity of estimating $\delta$ is inversely proportional to
$|\alpha|$. For convenience, we assume $n$ is even:
\begin{equation}
\langle cat_{n,\alpha}|cat_{n,\alpha}^{\delta}\rangle =
\frac{2}{n} \sum_{j=1}^{\frac{n}{2}} \cos\left[2 r \sin (\theta -
\frac{2 \pi j}{n})\right].
\end{equation}

It is easily checked that the OF, being of interferometric origin,
is only sensitive to the difference in phase : $\langle
cat_{n,\alpha}^{\delta_2}|cat_{n,\alpha}^{\delta_1}\rangle =
e^{\iota \phi}  \langle cat_{n,\alpha}|cat_{n \alpha}^{\delta_1 -
\delta_2}\rangle$. The OF for $n=2$,
\begin{equation}
|\langle cat_{2,\alpha}|cat_{2,\alpha}^{\delta}\rangle|^2 = \cos^2 (2 |\alpha|\delta_\perp)
\end{equation}
matches with the known result \cite{Toscano}, with \( \delta_\perp
= |\delta| \sin (\theta_\delta -\theta_\alpha) \text{ and }
\delta_\parallel = |\delta| \cos (\theta_\delta -\theta_\alpha)\).
As is depicted in Fig.~\ref{figure}, it is interesting to observe
that the above expression is analogous to the double slit
interference pattern, where the normalized intensity can be
written as $\frac{I}{I_{max}} = \cos^2 (\frac{y b \pi}{s
\lambda})$ \cite{Hecht}. The path difference between the two waves
reaching at the observation point is $yb/s$, where $b$ defines the
distance between the two slits, $s$ is the separation between the
aperture and the screen, and $y$ corresponds to the vertical
coordinate of the detector. The above analogy can be
mathematically established by taking $\lambda$ in the unit of $s$
and redefining the commutation relation, $[a,a^{\dagger}] = \pi
\lambda^{-1}$:
\begin{equation}
|\langle cat_{2,\alpha}|cat_{2,\alpha}^{\delta}\rangle|^2 = \cos^2
\left[2 \frac{|\alpha|\delta_\perp \pi}{\lambda}\right],
\end{equation}
where $2|\alpha|$ is the separation of the two coherent state
sources. Use of the phase shifted cat state,
$|cat2_{\alpha,\phi}\rangle = \frac{|\alpha\rangle+|\alpha
e^{\iota \phi}\rangle}{2}$, would yield an interference pattern at
an angle $\frac{\phi}{2}$ and \textit{fringe width},
$2|\alpha|\sin\frac{\phi}{2}$:
\begin{equation}
|\langle cat2_{\alpha,\phi}|cat2_{\alpha,\phi}^{\delta}\rangle|^2
=  \cos^2 (2 |\alpha|\sin\frac{\phi}{2} (\delta_\perp
\sin\frac{\phi}{2} + \delta_\parallel \cos\frac{\phi}{2}))
\end{equation}
Introducing a phase between the constituent CSs of a cat state
with $n=2$ gives the state $|cat2_{\alpha}^{\phi}\rangle =
\frac{|\alpha\rangle+e^{\iota \phi}|-\alpha\rangle}{2}$. The OF
for this state is
\begin{equation}
|\langle
cat2_{\alpha}^{\phi}|cat2_{\alpha}^{\phi,\delta}\rangle|^2 =
\cos^2 (2 |\alpha|\delta_\perp) - \phi) \text{,}
\end{equation}
akin to the phenomenon of ``\textit{fringe shift}'' observed in
classical optics.

We now derive the asymptotic limit to the state overlap and
sensitivity in parameter estimation,
\begin{eqnarray}
\varinjlim_{n \rightarrow \infty} \langle cat_{n,\alpha}|cat_{n,\alpha}^{\delta}\rangle
&=& \varinjlim_{n \rightarrow \infty} \frac{2}{n} \sum_{j=1}^{\frac{n}{2}} \cos (2 r \sin (\theta - \frac{2 \pi j}{n})) \nonumber   \\
&=& \varinjlim_{n \rightarrow \infty} \frac{1}{n} \sum_{j=1}^{n} \cos (2 r \sin (\theta - \frac{2 \pi j}{n})) \nonumber \\
&=& \int_0^1 \cos(2 r \sin (\theta - 2 \pi x)) dx \nonumber \\
&=& \frac{1}{2 \pi} \int_{0}^{2 \pi} \cos(2 r \sin (z)) dz \nonumber \\
&=& J_0(2 |\alpha| |\delta|) \text{.}\label{bessel}
\end{eqnarray}
This proves our assertion that states can be discriminated for
$|\delta| = \frac{C}{2 |\alpha|}$ due to orthogonality, where $C$
is a root of the Bessel function (of first kind) of order zero,
$i.e.,$ $J_0$.
\begin{figure}[htp]
\includegraphics[scale=0.5]{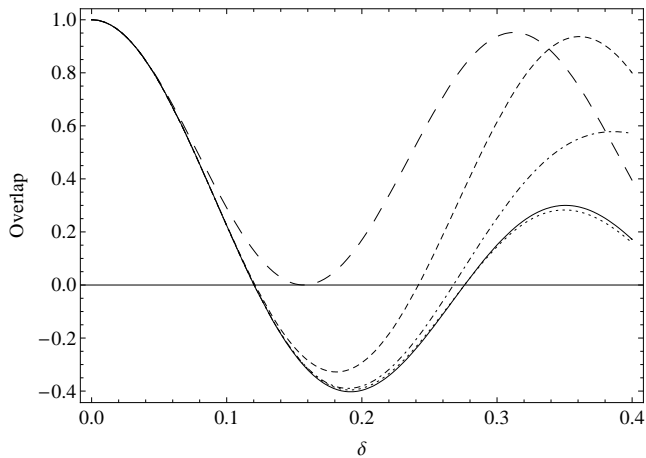}
\centering
\caption{Overlap function (Eq.~(\ref{over1}) or
(\ref{over2})) for superposition of four coherent states (dashed
line), superposition of six coherent states (small-dashed line),
superposition of eight coherent states (dot-dashed line), and
superposition of sixteen coherent states (dotted line). For larger
value of '$n$', the OF gradually coincides with the zeroth order
Bessel function (solid line). Here, $\alpha=10$ for all the
cases.} \label{overall}
\end{figure}

Although we have already discussed about the reliability of our
assumption before (see Fig.~\ref{offdiag}), we want to further
check, whether the above result is still valid for the general OF
in Eq.~(\ref{over1}) or (\ref{over2}). We provide
Fig.~\ref{overall} for numerical delineation, which clearly shows
the agreement of the zeroth order Bessel function for higher order
mesoscopic superposition. Hence, the experimental parameter domain
is well in the range of our previous assumption, where
Eq.~(\ref{bessel}) is valid.

The asymptotic overlap function [Eq.~(\ref{bessel})] is the result
of coherent superposition of $n-$CSs situated in a ring of radius
$\alpha$. For $n\rightarrow\infty$, the ring behaves as a ringed
shaped light source with constant angular source strength. Hence.
the superposition is analogous to the diffraction pattern
generated when light passes through the thin ring shaped opening.
The fact that the overlap between the cat states and their shifted
forms, is of the same form as the diffraction pattern centered at
one of the states, bears strong resemblance to the van
Cittert-Zernike theorem \cite{Wolf}, where the normalized mutual
coherence function $\gamma_{12}(0)$ between two points is
identical to diffraction pattern centered at one of the points.
For a ring shaped opening with constant angular source strength,
one can write explicitly
\begin{eqnarray}
 \gamma_{12}(0)&=&\frac{\langle  E_1(t) E_2(t)^{*}\rangle_T}{\sqrt{\langle E_1(t) E_1(t)^{*}\rangle_T \langle  E_2(t) E_2(t)^{*}\rangle_T}}\nonumber\\
&=& J_0\left(\frac{2 \pi r_{0} |\vec{r_1}-\vec{r_2}|}{\lambda
R}\right).
\end{eqnarray}
$\gamma_{12}(0)$ actually signifies the complex degree of spatial
coherence of the two points at the same instant in time, when
fields arriving at the observation screen being $E_1(t)$ and
$E_2(t)$ respectively. $r_{0}$ is the radius of ring, $R$ is the
distance of the screen from the opening and
$|\vec{r_1}-\vec{r_2}|$ is the path difference between the points.
The suffix $T$ in the expectation value signifies the time average
according to the ergodic hypothesis. The above equation should be
compared with the OF for large $n$ (Eq.~\ref{bessel}), for unit
distance from the screen to the opening ($R=1$) and for
$[a,a^{\dagger}] = \pi \lambda^{-1}$ :
\begin{equation}
\langle
cat_{n,\alpha}|cat_{n,\alpha}^{\delta}\rangle=J_0\left(\frac{2 \pi
|\alpha| |\delta|}{\lambda}\right).
\end{equation}
Thus the OF is similar to the mutual coherence function, nicely
matches with the diffraction pattern in accordance with the van
Cittert-Zernike theorem.

Considering the sinusoidal nature of the OF, it needs to be
checked where the states are not distinguishable. The fact that:
\begin{equation}
\langle cat_{n,\alpha}|cat_{n,\alpha}^{\delta}\rangle =
\frac{1}{n} \sum_{j=1}^n \sum_{k=1}^n \langle 0|D(e^{\frac{\iota 2
\pi j}{n} } \alpha )^{\dagger} D(\delta) D(e^{\frac{\iota 2 \pi
k}{n} } \alpha ) |0\rangle
\end{equation}
indicates that the maximum contribution of each term is
$\frac{1}{n}$ and minimum $\frac{-1}{n}$, which arise, if and only
if, either $D(\alpha)$ and $D(\delta)$ commute($\frac{1}{n}$) or
anti- commute($\frac{-1}{n}$). Thus, the state is
indistinguishable from its shifted form, if and only if,
displacement operators corresponding to all the constituent states
of the cat state commute with $D(\delta)$ or all anti- commute
with it.

\section{Conclusions}

In conclusion, the sensitivity of cat-like states to quantum
parameter estimation is studied for large number of constituent
CSs. The assumption of large phase space support is justified for
accessible parameter ranges in realistic situation. In this limit,
the state OF, determining the orthogonality of cat and displaced
cat states, approaches the Bessel function. According to the van
Cittert-Zernike theorem, the coherence problem is mathematically
identical with the diffraction problem by complex degree of
coherence. The fact that the OF is having the same form as the
diffraction pattern results the same expression of normalized
mutual coherence function for large $n$. This is similar to the
mutual coherence function of a circular ring, which yields Bessel
function of order zero, matching with the theorem of van
Cittert-Zernike.

\section{Acknowledgment}

This work was supported by the National Initiative on
Undergraduate Science (NIUS) undertaken by the Homi Bhabha Centre
for Science Education-Tata Institute of Fundamental Research
(HBCSE-TIFR), Mumbai, India. Discussions with Prof. Vijay. A.
Singh of HBCSE are gratefully acknowledged.

\end{document}